\documentclass[12pt]{article}

\usepackage{amsmath,amssymb,amsfonts,textcomp}
\usepackage{hyperref}
\usepackage{color}
\usepackage{float}
\usepackage{rotating}
\usepackage[round]{natbib}
\begin{document}

\title{Material Facts Obscured in Hansen's Modern Gauss-Markov Theorem}
\author{H D Vinod
\thanks{Email  vinod@fordham.edu.}
}

\maketitle 

\begin{abstract}
We show that the abstract and conclusion of Hansen's {\it Econometrica} paper, \cite{Hansen22}, entitled a
modern Gauss-Markov theorem (MGMT),   
obscures a material fact, which in turn can
confuse students. The MGMT places
ordinary least squares (OLS) back on a high pedestal by
bringing in the Cramer-Rao efficiency bound. We explain
why linearity and unbiasedness are linked, making most
nonlinear estimators biased.  Hence,
MGMT 
extends the reach of the century-old GMT by a near-empty set.
It misleads students because it misdirects attention back to 
the unbiased OLS from beneficial
shrinkage and other tools, which reduce the mean squared error
(MSE) by injecting bias.
\end{abstract}

\section{Introduction}
\label{sec.intro}

The century-old Gauss-Markov theorem (GMT)
showed that the ordinary least squares (OLS) estimator (sample mean)
is the (minimum variance) best (efficient) linear unbiased estimator (BLUE). 
 \cite{Stein55} shocked statisticians when he
proved the inadmissibility of OLS, and
threw
the first
salvo toward the break-up of the
long-dominating GMT.
Numerous subsequent papers established that the 
so-called Stein-rule (nonlinear shrinkage) estimator is
guaranteed to reduce the mean squared error (MSE) of OLS.
\cite{EfronMorris73} explain
the practical significance of Stein-rule in
a {\it Scientific American}  magazine article (May 1977)
using baseball batting averages.  The key point of the
Stein-rule is that despite the GMT BLUE result,
shrinkage estimation beats OLS.

Ridge regression is another strand of literature rejecting
OLS. It is also a biased nonlinear shrinkage estimator.
\cite{HDVSignif} explains the ``big idea'' behind
ridge regression, and
concludes with a quote from \cite{Hastie20}:
``What started out as a simple fix for wayward linear regression
models has evolved into a large collection of tools for data modeling. It would
be hard to imagine the life of a data scientist without them.''

With considerable ingenuity,
\cite{Hansen22} claims to
generalize GMT, calling it modern. He 
drops the modifier  “L=linear”
in BLUE. He claims to improve upon other efforts 
by Kariya, Kurata, Berk, and Hwang to generalize
GMT. Others are too
restrictive on the class of allowed nonlinearity
or error distributions.
The abstract and conclusion sections
of \cite{Hansen22}
give the false impression that OLS remains best even
in a class of nonlinear estimators. 
However, the eleven pages of {\it Econometrica} 
devoted to MGMT do not include any 
Efron-Morris type example of a nonlinear unbiased
estimator.

Hansen fails to reveal that the adjectives
linear and unbiased (L\&U) are like conjoint twins in the
present context, since all
nonlinear estimators used in practice are biased in finite samples. 
Hence, MGMT dropping the modifier L while keeping its twin U
(unbiased) is primarily an empty change of little practical
consequence. 
The gap between BLUE and BUE is hairLine-thin. I understand that
group theory in pure mathematics has  translation-invariant
groups. These `groups' can yield esoteric  
quadratic functions to design
an unbiased nonlinear estimator. The exception 
proves the rule that
the class of nonlinear unbiased estimators
is not simply ``small," as Hansen admits, but near empty.

\section{Linearity \& Unbiasedness Conjoint twins}
Now we explain why it
is hard to separate the conjoint twins (L\&U). We
begin with 
the linear regression model with $p$ regressors. In standard notation, it is
$y=X\beta +\epsilon$. The
OLS estimator is
$b=(X^\prime X)^{-1}X^\prime y$.
The linearity of the OLS model $y=X\beta +\epsilon$ allows us to write
\begin{equation}
b=(X^\prime X)^{-1}X^\prime y = (X^\prime X)^{-1}X^\prime X\beta + \epsilon
\end{equation}
where $b$ and $\beta$ are $p\times 1$ vectors. Now compute the
expectation of $b$ using  the assumption that $E\epsilon=0$.  We have
$Eb=\beta$, where $(X^\prime X)^{-1}(X^\prime X)=I$.  
Also, OLS residuals $\hat \epsilon_t$ add up to zero in finite samples,  $\Sigma_t\hat \epsilon_t=0$.

It is important to note that
one can compute the expectation of each regression coefficient 
separately, only because of the linearity of the OLS model.
The separation difficulty creates the conjoint twins of linearity and unbiasedness (L\&U)
in the context of individual coefficient estimates in finite samples.

A corresponding
nonlinear model is $y_t = f(X\beta)_t + u_t$, where the nonlinear
regression function $f(X\beta)_t$ depends on a vector of
coefficients $\beta$. Unlike linear
models, components of $\beta$ can include complicated functions,
including various powers and trigonometric transformations.
Unbiasedness requires proof that each estimated nonlinear coefficient
based on finite samples
be equal to its true population value, 
$E(\hat \beta) = \beta$.

Since there is no
commonly accepted method for estimating the nonlinear parameters
$\hat \beta$, which can be sensitive to starting values,
one cannot simply write the expression
for the expectation of a single component $\beta_j$ which
does not also depend on other components $\beta_k,  k\ne j$. 
Clean expressions for only one
nonlinear coefficient at a time may not exist. 
It is thus established that most individual
nonlinear estimators are biased, $E(\hat \beta) \ne \beta$.
Hence, our claim of conjoint twins (L\&U) is
supported.

Consider a textbook example from the R code
snippet \#R1.2.1 in \cite{Vinod:B08}. Let
$y$ denote output, $K$ denotes capital input, and $L$ denotes labor input. A nonlinear Cobb-Douglas production function is
\begin{equation}
	\label{eq.cobbD}
	y=A K^\alpha L^\beta + \epsilon.
\end{equation}

The exact implication of 
Hansen's modern GMT to the nonlinear least squares (NLS) estimator
of (\ref{eq.cobbD}) is unclear.  The textbook
snippet uses metals data
to estimate NLS coefficients
$(A, \alpha, \beta)$.  
If one plugs in the NLS estimates to define
the fitted value of output $\hat y$, the residuals,
$\hat \epsilon =y-\hat y$, do not add up to zero, or
$\Sigma \hat\epsilon \ne 0$.  Then,
least squares estimates $(\hat A, \hat\alpha, \hat \beta)$
must be biased.  Hence, Hansen's modern GMT does not apply to the estimation of Cobb-Douglas nonlinear functions.

\section{Final Remarks}
Hansen's conclusion states, ``The Gauss-Markov Theorem is a core efficiency result 
but restricts attention to linear estimators-–-and this is an inherently uninteresting restriction."
We agree. However, Hansen fails to disclose that his
MGMT does not at all cover most of the interesting
nonlinear estimators used in Econometrics.
Hansen's MGMT
has not reestablished the efficiency of OLS, except for an
uninteresting, near-empty
set of nonlinear unbiased estimators.

The abstract and conclusion of a research paper are
akin to advertisements.  A
Federal Trade Commission (FTC) employee,
\cite{FTCFullDisclosure}, formally 
defines what is expected from advertisers. They must disclose all
``material'' facts about a product,
so that they satisfy a mnemonic of four P's: 
(i) {\bf P}rominent display, (ii) understandable  {\bf P}resentation, 
(iii) a visible {\bf P}lacement, and (iv) a {\bf P}roximity to the claim
being modified.  
This note shows that \cite{Hansen22} fails to satisfy these norms.

Hansen is careful to include 
a caveat in an obscure part of his paper stating
that ``the class of nonlinear unbiased estimators is small.''
The caveat fails to clarify that
the relevant set is near empty, not just small. 
More important, the caveat is ``material" to
the main point of MGMT. In the interest of ethical
disclosure norms, the caveat should have been  more
conspicuously
presented, proximate to the claims in the abstract and conclusion
of Hansen's paper.  Otherwise,  readers may
be tempted to conclude that OLS beats many nonlinear
estimators, a patently false impression.

Instead of citing the more relevant {\cite{Stein55}},
Hansen cites the insight by Stein (1956). This
should not give a false
impression that Hansen has co-opted the Stein-rule.
This note
hopes to explain the presence of conjoint twins
(L\&U) in regression model estimation from
finite samples. It is hard to separate linearity from
unbiasedness. Hence Hansen's MGMT extends the set
of estimators where OLS is more efficient than nonlinear
estimators
by an uninteresting, nearly empty set.
A generalization that extends the applicability
of GMT by one esoteric unbiased nonlinear estimator is
not a worthy generalization.
\bibliographystyle{asa}
\bibliography{c:/mybib/ref12}

\end{document}